\documentclass[aps,twocolumn,superscriptaddress,prl,showpacs,bibnotes]{revtex4}
\usepackage{graphicx}
\usepackage[usenames]{color}

\begin{document}


\title{Long exciton spin memory in coupled quantum wells}



\author{K.~Kowalik-Seidl}\email[Corresponding author: ]{k.kowalik@physik.uni-muenchen.de}

\author{X.~P.~V\"{o}gele}

\author{B.~N.~Rimpfl}

\author{S.~Manus}\affiliation{Center for NanoScience and Fakult\"{a}t f\"{u}r Physik, Ludwig-Maximilians-Universit\"{a}t, Geschwister-Scholl-Platz 1,
D-80539 M\"{u}nchen, Germany}

\author{D.~Schuh}

\author{W.~Wegscheider}\affiliation{Institut f\"{u}r Experimentelle und Angewandte Physik, Universit\"{a}t Regensburg, D-93040 Regensburg, Germany}

\author{A.~W.~Holleitner}\affiliation{Walter Schottky Institut and Physik Department, Technische Universit\"{a}t M\"{u}nchen, D-85748 Garching, Germany}

\author{J.~P.~Kotthaus}\affiliation{Center for NanoScience and Fakult\"{a}t f\"{u}r Physik, Ludwig-Maximilians-Universit\"{a}t, Geschwister-Scholl-Platz 1,
D-80539 M\"{u}nchen, Germany}


\date{\today}

\begin{abstract}
Spatially indirect excitons in a coupled quantum well structure were studied by means of polarization and time resolved photoluminescence. A strong degree of circular polarization ($>50\%$) in emission was achieved when the excitation energy was tuned into resonance with the direct exciton state. The indirect transition
remained polarized several tens of nanoseconds after the pumping laser pulse, demonstrating directly a very long relaxation time of exciton spin. The observed spin memory effect exceeds the radiative lifetime of the indirect excitons.

\end{abstract}
\pacs{78.47.jd,  78.67.De}

\maketitle

\indent Semiconductor heterostructures with tunable spatial separation of electrons and holes attract a great attention for their possible applications in
optoelectronic devices~\cite{Krauss-photonicImages, High-ExcIntegratedCircuits, High-OptoelTransistor, Snoke-ExcCircuits}. While the storage of excitons for
several microseconds has been experimentally evidenced~\cite{Krauss-photonicImages,Rocke-AcoustStorage, Zimmermann-PhotStorage, Lundstrom-ExcStorage,
Winbow-photStorageCQW}, it is still an open question how to exploit the spin of indirect excitons for information storage. Recent experiments performed on coupled quantum wells (CQWs) reported exciton spin transport in the range of micrometers, suggesting indirectly a long exciton spin relaxation time of nanoseconds~\cite{Leonard-SpinTransportExc}. These results suggest that a local probe can overcome the spin relaxation caused by the spin propagation dynamics. Here, we exploit a confocal optical scheme in order to study the spin dynamics of indirect excitons at the location of their excitation. Hereby, we resolve the spin memory of indirect excitons in coupled quantum wells via time- and polarization-resolved PL studies after resonant and non-resonant excitation of the direct and indirect exciton states. We observe that under conditions of resonant excitation a highly efficient initialization of exciton spin takes place. The PL remains strongly circularly polarized long after the laser pulse and nearly constant during the lifetime of excitons. Our results directly confirm a long spin relaxation time of $> 80$ nanoseconds for indirect excitons. This time scale is an order of magnitude longer than the one obtained by read-out schemes with a larger focus spot ~\cite{Larionov-SpinRelaxDQW}.

\begin{figure}[h]
\begin{center}
\includegraphics[width=0.5\textwidth,keepaspectratio]{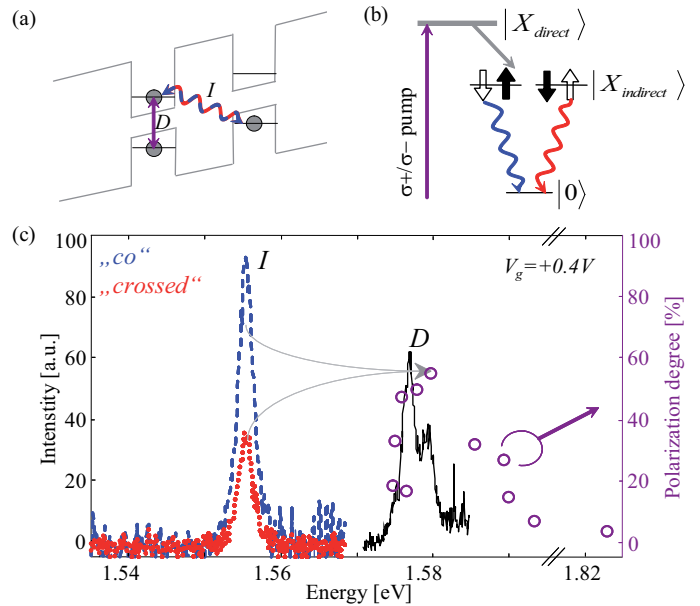}%
\end{center}
\caption{(Color online) (a) Scheme of GaAs CQW under applied electric field showing formation of direct (\emph{D}) and indirect (\emph{I}) excitons. (b) Diagram of
resonant polarized excitation in direct excitonic states and emission from two optically active indirect states. (c) Left scale: PL spectra for direct excitons (\emph{D}, solid black line) under non-resonant excitation and for indirect excitons (\emph{I}) under resonant $1.58$~eV excitation, polarization resolved in "co" (dashed blue) and "cross" (dotted red) configurations. Right scale: degree of circular polarization for indirect excitons vs. excitation laser energy (open circles). The gray arrows indicate the point obtained from the spectra shown in the left part of the figure.}\label{fig:1}
\end{figure}

\indent The studied heterostructure consists of two 8-nm wide GaAs/AlGaAs coupled quantum wells separated by a 4-nm Al$_{0.3}$Ga$_{0.7}$As barrier and it was fabricated as a field-effect device~\cite{Gaertner-drift, Gaertner-Si0trapsSample, Voegele-PRL}. The shape of the confining potential is controllably adjusted by the bias ($V_{g}$) applied between a semitransparent metal gate and a deep ohmic contact, allowing the direct manipulation of the exciton lifetime via a modified electron-hole separation (Fig.~\ref{fig:1} (a)). The PL excitation and collection are performed through the semitransparent part of the top Schottky gate using a confocal microscope based on a short focal length aspheric lens. The luminescence is dispersed by a 0.5~m double monochromator and detected with an intensified charge coupled device (ICCD) detector. The configuration of the setup allows us to perform measurements under identical ("co") and orthogonal ("cross") polarizations of excitation and detection. The resulting extinction ratio is below $0.0015$ for the two opposing circular polarizations. All measurements are performed at 4.2~K.

\indent In order to precisely define the initial spin polarization of excitons we optimize the process of resonant excitation. In CQWs the bright excitons form a two level system and the eigenstates $|\pm 1 \rangle$ are coupled to circularly polarized photons (see scheme in Fig.~\ref{fig:1}(b)). As a strictly resonant excitation and PL measurements are rather incompatible, we take advantage of the quasi-resonant excitation into spatially direct exciton states, which has proven to be very efficient in similar structures~\cite{Leonard-SpinTransportExc}. Such a quasi-resonant and circularly polarized excitation should provide an efficient transfer of light polarization into exciton spin polarization. The two quantum wells in the heterostructure are identical. Therefore a quasi resonant excitation results in pumping simultaneously the direct transitions in both wells. Since the direct PL recombination lifetime is typically $1$~ns~\cite{Feldmann-DirectLifetime}, the conversion of direct into indirect states is likely related with a fast carrier tunneling between the wells, which is of the order of tens of picoseconds (estimated from WKB approximation~\cite{Tunneling}). Then, the exciton recombination transforms the spin polarization of the tunneled charged carriers into the polarization of the emitted photon. Thus, the optical polarization of the indirect excitons provides a direct and very sensitive tool to study spin memory effect in coupled quantum wells (Fig.~\ref{fig:1}(b)).

\indent First we employ a $1.823$~eV ($680$~nm) laser diode to examine a non-resonant excitation. The scheme is used to measure the PL of the direct excitons and to determine the optimal excitation conditions for indirect emission (black line in Fig.~\ref{fig:1} (c)). Then, the PL from indirect excitons is measured for different voltages and under "co" and "cross" configurations of the circular polarization in the excitation and the detection paths. First a continues wave laser diode is used in the experiment. Tuning systematically the laser diode energy in the range $1.57-1.59$~eV ($780-790$~nm) enables us to drive the excitation through the direct excitonic states. We measure the degree of circular polarization of indirect excitons $P_{circ} = \frac{I_{co}-I_{cross}}{I_{co}+I_{cross}}$ as a normalized difference between PL intensities $I_{co}$ and $I_{cross}$ in "co" and "cross" configurations, respectively. Figure~\ref{fig:1}(c) shows the two spectra of indirect excitons taken in "co" and "cross" polarization configurations under strictly resonant excitation ($1.58$~eV) into the direct excitons. The right hand side of the figure illustrates the degree of circular polarization for indirect excitons as a function of excitation laser energy. For non-resonant excitation at $1.82$~eV no significant polarization of indirect excitons PL is measured, whereas a clear enhancement of the circular polarization is observed for a quasi-resonant excitation of the direct excitonic states. The maximum value of $P_{circ}$ is about $55$~\% for the excitation at $1.58$~eV. Such efficient polarization conservation suggests a long relaxation time of the exciton spin, which exceeds the radiative recombination time.

\indent The PL circular polarization can be written as~\cite{Dzhioev-conversion, Optical-Orientation}:

\begin{equation}
P_{circ}= \frac{\tau_s}{\tau_s+\tau_r} P_{circ}^0 \label{eq:CircPolar}
\end{equation}

\noindent where $\tau_s$ ($\tau_r$) is the spin relaxation time (radiative lifetime) of excitons and $P_{circ}^0$ is the effective initial circular polarization determined by the laser polarization and losses during the formation of an indirect exciton. Hereby, we estimate that the spin relaxation under strictly resonant excitation is at least $1.2$ times longer than the exciton lifetime, when neglecting any polarization loss in the excitation process $P_{circ}^0=1$. The radiative lifetime for indirect excitons reaches several tens of nanoseconds, thus the spin relaxation time must be even longer.

\begin{figure}[h]
\begin{center}
\includegraphics[width=0.5\textwidth,keepaspectratio]{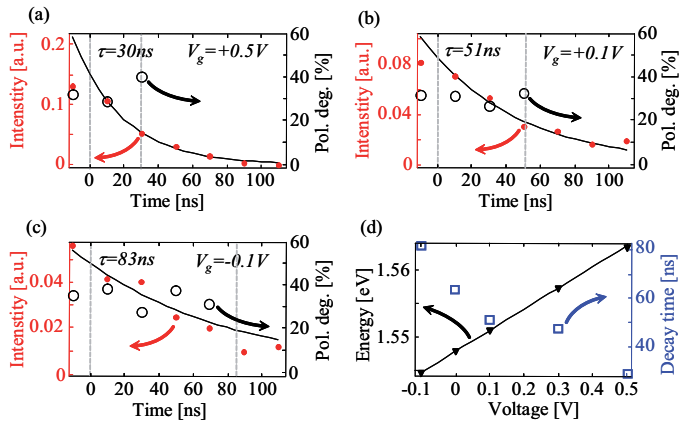}%
\end{center}
\caption{(Color online) (a)-(c) For 3 different applied biases under $1.577$~eV excitation: left scale -- PL decay vs. delay time after the excitation
pulse (filled circles -- experimental data, line -- exponential fit), right scale -- circular polarization degree (open circles). Vertical dashed lines mark zero delay and the one corresponding to the PL decay time, respectively. (d) Dependence of transition energy on applied bias (black triangles -- experimental data, line -- linear fit of the Stark shift), and of PL decay time (opened blue squares).}\label{fig:2}
\end{figure}

\indent To confirm straightforwardly this hypothesis we apply a pulsed excitation at $1.577$~eV and "co" and "cross" polarized photoluminescence spectra are recorded for different delays after the $400$~ns long laser pulse~\cite{NegativeDelay}. Figure~\ref{fig:2}(a)-(c) presents the time evolution of the PL after the laser pulse (filled circles), measured for different gate voltages. The emission decay rate, obtained from the exponential fit~\cite{NegativeDelay}, shows its increase with applied electric field~\cite{Alexandrou-QWelecField} as pictured in Fig.~\ref{fig:2}(d). The obtained circular polarization degree $P_{circ}$ is plotted up to delay times comparable with radiative decay time in Figure~\ref{fig:2}(a)-(c). The degree of circular polarization for the indirect exciton emission remains constant several tens of nanosecond after the laser pulse. For an applied bias of $-0.1V$~V the radiative lifetime is measured to be as long as $83$~ns and PL polarization is constantly $\sim 35$~\% during this period (see Figure~\ref{fig:2}(c)).

\indent Generally, the spin polarization of excitons is maintained only for very short times in a single quantum well. The spin relaxation is dominated by dephasing phenomena which results from carriers exchange interaction~\cite{Maialle-SpinDynSingleQW, Vinattieri-ExcDynSQW, Urdanivia-ExcDynSQW}. The large electron-hole separation in CQWs suppresses this interaction and enables a long relaxation time for excitonic spin. We note that $\tau_s$ in CQW obtained from our studies exceeds few times the one estimated from the other experiments~\cite{Larionov-SpinRelaxDQW, Leonard-SpinTransportExc}. This difference might come from the different applied experimental methods. In Ref.~\cite{Larionov-SpinRelaxDQW} only very short $<1$~ns relaxation times were measured, but the authors noted that high-energy part of the spectrum conserved the exciton spin for much longer time scales (few ns) than the lower energy tail. As their signal was investigated only in the time domain, but without confocal scheme, it averaged over big area of drifting excitons. The authors of Ref.~\cite{Leonard-SpinTransportExc} studied PL in spatial domain and they used the exciton diffusion to estimate $\tau_s$ from the PL polarization in continuous wave experiment. Their indirect method gave $\tau_s \sim 10$~ns. Our technique allowes to directly measure the time evolution of the spin polarization. In our confocal microscope the emission and collection aperture is limited to $\sim1\mu m$. Hence, we would like to emphasize that only excitons created at the excitation spot are investigated in our experiment. Likely the losses of the spin polarization during the exciton drift induced the reduction of $\tau_s$ in Refs.~\cite{Larionov-SpinRelaxDQW, Leonard-SpinTransportExc} in comparison to our results.

\indent To conclude, the polarization and time resolved PL of the indirect emission from coupled GaAs/AlGaAs quantum wells is investigated. Quasi-resonant excitation is successfully employed to initialize the exciton spin population via the direct excitons and a subsequent spin transfer and spin storage into indirect excitons. Our findings from continuous and time-resolved experiments directly demonstrate a long spin relaxation time of indirect excitons in CQWs exceeding $80$ nanoseconds.

\indent The authors gratefully acknowledge helpful discussions with A. O. Govorov. One of us (K.~K.-S.) is supported by the Alexander von Humboldt Foundation. We
acknowledge financial support by the Center for NanoScience (CeNS), the Nanosystems Initiative Munich (NIM), LMUexcellent, the DFG Project KO 416/17 and HO 3324/4.



\begin{thebibliography}{22}
\bibitem[{Neg()}]{NegativeDelay}
\bibinfo{note}{The generated laser pulses have rectangular shape with fall and
  rise times of $\sim1$~ns. The "zero" of the time scale is defined by the end
  of the laser pulse. Negative values correspond to the measurements taken
  under laser illumination. For the estimation of the decay time we take only
  experimental points recorded after the laser pulse.}
\end{thebibliography}


\end{document}